\documentclass[prl,reprint,superscriptaddress]{revtex4-1}

\usepackage{amsmath,amssymb}
\usepackage{verbatim}
\usepackage{graphicx}
\usepackage{hyperref}
\usepackage{color}
\usepackage{mathrsfs}

\newcommand {\cL}{{\cal L}}

\newcommand {\cW}{{\cal W}}

\def\a{\alpha}
\def\b{\beta}

\def\e{\epsilon}

\def\l{\lambda}
\def\m{\mu}
\def\n{\nu}
\def\o{\omega}
\def\p{\pi}

\def\r{\rho}
\def\s{\sigma}

\def\ri{{\rm i}}

\newcommand{\pa}{\partial}

\newcommand{\be}{\begin{equation}}
	\newcommand{\ee}{\end{equation}}
\newcommand{\bea}{\begin{eqnarray}}
	\newcommand{\eea}{\end{eqnarray}}
\newcommand{\non}{\nonumber}
\newcommand{\ba}{\begin{array}}
	\newcommand{\ea}{\end{array}}

\def\double #1{#1{\hbox{\kern-2pt $#1$}}}

\newcommand{\bsubeq}{\begin{subequations}}
	\newcommand{\esubeq}{\end{subequations}}

\newcommand{\de}{{\nabla}}

\newcommand{\nn}{\nonumber}

\def\ft#1#2{{\textstyle{\frac{\scriptstyle #1}{\scriptstyle #2} } }}

\def\rmi{{\rm i}}

\def\a{\alpha}
\def\b{\beta}

\def\e{\epsilon}

\def\p{\psi}

\def\l{\lambda}

\def\m{\mu}
\def\n{\nu}
\def\r{\rho}
\def\s{\sigma}

\def\o{\omega}

\begin{document}

	\title{All Gauged Curvature Squared Supergravities in Five Dimensions}
	
	\author{Gregory Gold}
	\email{g.gold@uq.edu.au}
	\affiliation{School of Mathematics and Physics, University of Queensland, 
		St Lucia, Brisbane, Queensland 4072, Australia}
	
	\author{Jessica Hutomo}
	\email{j.hutomo@uq.edu.au}
	\affiliation{School of Mathematics and Physics, University of Queensland, 
		St Lucia, Brisbane, Queensland 4072, Australia}
	
	\author{Saurish Khandelwal}
	\email{s.khandelwal@uq.edu.au}
	\affiliation{School of Mathematics and Physics, University of Queensland, 
		St Lucia, Brisbane, Queensland 4072, Australia}
	
	\author{Mehmet Ozkan}
	\email{ozkanmehm@itu.edu.tr}
	\affiliation{Department of Physics,
		Istanbul Technical University,
		Maslak 34469 Istanbul,
		Turkey}
	
	\author{Yi Pang}
	\email{pangyi1@tju.edu.cn}
	\affiliation{Center for Joint Quantum Studies and Department of Physics,\\
		School of Science, Tianjin University, Tianjin 300350, China \\}
	
	\author{Gabriele Tartaglino-Mazzucchelli}
	\email{g.tartaglino-mazzucchelli@uq.edu.au}
	\affiliation{School of Mathematics and Physics, University of Queensland, 
		St Lucia, Brisbane, Queensland 4072, Australia}

	\date{\today}

	\begin{abstract}
		
We present  a complete basis to study gauged curvature-squared supergravity in five dimensions.   We replace the conventional   ungauged Riemann-squared action with a new Log-invariant, offering a comprehensive framework for all gauged curvature-squared supergravities. Our findings address long-standing challenges and have implications for precision tests in the AdS/CFT correspondence.
		
	\end{abstract}
	
 	\maketitle
	\allowdisplaybreaks

\textit{Introduction.}---Twenty-six years after its discovery, the AdS/CFT correspondence has entered a new era in which precision tests beyond the leading order have become increasingly important, owing to developments in both field theory and gravity. On the one hand, integrability and localization techniques allow one to compute the observables in superconformal field theories (SCFT) exactly at finite couplings. On the other hand, the development of superconformal tensor calculus and superspace techniques -- see the reviews \cite{Freedman:2012zz,Lauria:2020rhc,Kuzenko:2022skv,Kuzenko:2022ajd} -- in conjunction with the computational capabilities offered by computer algebra programs, has significantly advanced the construction of exact, off-shell higher-derivative supergravity models.

 In this letter, we present all gauged curvature-squared supergravity invariants in five dimensions based on the  off-shell dilaton Weyl multiplet. After going on-shell, our invariants describe the most general four-derivative corrections to the five-dimensional minimal gauged supergravity, which is a universal sector to all string compactifications preserving at least eight supercharges. The gauged aspect is necessary to accommodate a supersymmetric anti-de Sitter (AdS) solution, and thus is of broad interest in holography. In particular, due to recent advancements in AdS black hole microstate counting \cite{Benini:2015eyy,Benini:2016rke,Cabo-Bizet:2018ehj,Choi:2018hmj,Benini:2018ywd,Honda:2019cio,Benini:2020gjh,Agarwal:2020zwm} using the dual CFT, a precise matching between the gravity and CFT results at the next to leading order clearly requires the knowledge of the complete curvature-squared supergravity actions. Previous works have made attempts to compute four-derivative corrections based on partial results in the literature, and certain assumptions were made. Using our full results, it can be shown that, in fact, some of the assumptions are invalid, thus finally furnishing the stage for new, next-to-leading order analyses on the gravity side of the AdS/CFT correspondence.

 The construction of gauged curvature-squared invariants is notoriously hard, as opposed to their ungauged counterparts, which have been fully known for more than a decade \cite{Bergshoeff:2011xn,Ozkan:2013nwa,Ozkan:2013uk}. The primary difficulty stems from the absence of a straightforward transition from ungauged to gauged theories. In fact, the complete basis of invariants must be constructed from completely different starting points. For instance, certain ungauged curvature-squared models are attainable through the application of superconformal tensor calculus, utilizing the dilaton Weyl multiplet. In contrast, their gauged counterparts necessitate the use of a modified version of the same multiplet \cite{Coomans:2012cf}, which has an entirely different field content and transformation rules. Furthermore, the deformation necessary for the construction of gauged supergravity models renders certain established higher-derivative supergravity building techniques impractical, thus further complicating the task.  In fact, it takes an interplay between superconformal tensor calculus  \cite{Coomans:2012cf} and superspace techniques \cite{Butter:2014xxa}, together with a series of new, daunting computations finalized only in the results presented here, to yield the complete set of gauged curvature-squared invariants.

This letter aims to explicitly show that the past challenges can be overcome by changing the basis of curvature-squared supergravities, which previously employed the Weyl tensor squared, Riemann tensor squared, and Ricci scalar squared as fundamental building blocks. We demonstrate that by replacing the Riemann-squared action with the Log-invariant, in which the leading term comes with the Ricci tensor squared, it is possible to explicitly establish all gauged curvature-squared supergravities in five dimensions. The outcomes presented in our letter mark a significant advancement, paving the way to a complete study of physical results beyond the leading supergravity approximation in five dimensions. This development holds particular promise for precision tests of the AdS$_5$/CFT$_4$ correspondence. In this context, we derive the anomaly coefficients in the dual SCFT$_4$, which apparently depend on all curvature-squared couplings.

\textit{Construction of the invariants.}---We start by introducing the field content of the standard Weyl multiplet of conformal supergravity in five dimensions \cite{Bergshoeff:2001hc}. Our notation and conventions correspond to that of \cite{Bergshoeff:2001hc}. We denote the spacetime indices by $\mu, \nu, \cdots$, Lorentz indices by $a, b, \cdots$, SU(2)  indices by $i, j, \cdots$, and spinor indices by $\a, \b, \cdots$. The multiplet is described by a set of independent gauge fields: the vielbein $e_\mu{}^a$,	the gravitino $\psi_{\mu}{}_{\a}^i$,	the SU(2) gauge fields $V_\mu{}^{ij}$, and a dilatation gauge field $b_\mu$. The other gauge fields associated with the remaining symmetries, including the spin connection  $\omega_\mu{}^{a b}$, the $S$-supersymmetry connection	$\phi_{\m}{}_{\a}^i$, and the special conformal connection	${f}_\mu{}^a$, are composite fields, i.e., they are determined in terms of the other fields by imposing certain curvature constraints. The standard Weyl multiplet also contains a set of matter fields: a real antisymmetric	tensor $T_{a b}$, a fermion $\chi_\a^i$, and a real scalar $D$. A more detailed discussion of the superconformal transformations of the various fields can be found, e.g., in \cite{Bergshoeff:2001hc,Coomans:2012cf}.

	Below we will make use of a variant multiplet of conformal supergravity, known as the gauged dilaton Weyl multiplet \cite{Coomans:2012cf,Note2}. For this multiplet, the independent gauge fields remain the same as the standard Weyl multiplet, but the matter content is replaced with $\{\sigma, C_\mu, B_{\mu\nu}, L_{ij}, E_{\m\n\r}, N, \psi^i, \varphi^i\}$. This is obtained by coupling the standard Weyl multiplet to on-shell vector and linear multiplets. The vector multiplet consists of a scalar field $\s$, the gaugino $\psi_{\a}^i$, an abelian gauge vector $C_{\mu}$ with field strength $G_{\mu \nu}= 2 \pa_{[\mu} C_{\nu]}$, and an $\rm{SU(2)}$ triplet of auxiliary fields $Y^{ij} = Y^{(ij)}$. The linear multiplet contains an $\rm{SU(2)}$ triplet of scalars $L^{ij} = L^{(ij)}$, a gauge three-form $E_{\m\n\r}$, a scalar $N$, and an $\rm{SU(2)}$ doublet $\varphi^i_{\alpha}$. The bosonic matter fields of the vector and the standard Weyl multiplet are then expressed as follows \cite{Coomans:2012cf}
	\bea 
	Y^{ij} &=&   - \ft{g}{2} \s^{-1} L^{ij} + {\rm f.t.}~, 
 \nn\\
	T_{ab} &=& \ft{1}{8}\sigma^{-1} G_{ab} + \ft{1}{48} \sigma^{-2} \e_{abcde} H^{cde} + {\rm f.t.}~, 
 \nn\\
	D &=&  \ft{1}{4} \sigma^{-1} \nabla^a \nabla_a \sigma + \ft{1}{8} \sigma^{-2} (\nabla^a \sigma) \nabla_a \sigma -\ft{1}{32} R
 \non\\
	&& - \ft{1}{16} \sigma^{-2} G^{ab} G_{ab} - ( \ft{26}{3}  T^{ab} -2 \sigma^{-1} G^{ab} )  T_{ab}
	\non\\
	&& + \ft{g}{4} \s^{-2} N   + \ft{g^2}{16} \s^{-4} L^2 + {\rm f.t.}~,
 \label{dilaton weyl basis}
	\eea
	where ``f.t." stands for omitted fermionic terms and $H_{abc} = e_{a}{}^{\mu} e_{b}{}^{\nu} e_{c}{}^{\rho} H_{\mu \nu \rho}$ denotes the three-form field strength $H_{\mu \nu \rho}:= 3 \partial_{[\mu}B_{\nu \rho]}+\frac{3}{2}C_{[\mu}G_{\nu \rho]} + \frac12 g  E_{\m\n\r}$.	In the above, the covariant derivative is denoted by
	\bea
	\nabla_a = e_{a}{}^{\mu} \big( \pa_{\mu}-  \o_{\mu}{}^{b c} M_{bc}- b_{\mu} \mathbb{D} - V_{\m }{}^{i j} U_{i j} \big)~,
	\eea
	with $M_{ab}$, $\mathbb{D}$, and $U_{ij}$ being the Lorentz, dilatation, and SU(2) generators, respectively. The dilatation connection $b_{\mu}$ is pure gauge and will be set to zero throughout. The mapping \eqref{dilaton weyl basis} allows us to easily convert every invariant involving a coupling to the standard Weyl multiplet to that written in terms of the gauged dilaton Weyl multiplet. The ungauged map and the models can simply be obtained by setting $g=0$ in \eqref{dilaton weyl basis}. In this case, the fields of the linear multiplet decouple from the map \eqref{dilaton weyl basis}, and the multiplet reduces to the ungauged dilaton Weyl multiplet with $32+32$ off-shell degrees of freedom \cite{Bergshoeff:2001hc, Fujita:2001kv}.
 	
In the superconformal tensor calculus, the so-called BF action principle plays a fundamental role in the construction of general supergravity-matter couplings, see \cite{Butter:2014xxa,Kugo:2000af,Fujita:2001kv,Kugo:2002vc,Bergshoeff:2001hc,Bergshoeff:2002qk,Bergshoeff:2004kh} for the 5D case. It is based on an appropriate product of a linear multiplet with an Abelian vector multiplet:
	\bea
	&&e^{-1} \cL_{\rm BF} = \,A_{{a}} E^{{a}} + \rho N + \mathcal{Y}_{i j} L^{i j} + \rm f.t.~.
	\label{BF-Scomp}
	\eea
	Here we use $\{\rho, A_\m, \mathcal{Y}_{ij}, \lambda^i_\alpha\}$ to denote the field content in an arbitrary vector multiplet, and the bosonic part of the constrained vector $E_a$ is related to the three-form gauge field $E_{abc}$ via $E_a = -\frac{1}{12} \e_{abcde} \nabla^b E^{cde}$. In any construction that involves composite expressions for the fields of the linear multiplet in terms of the vector multiplet, the BF-action yields a vector-coupled action in the (gauged) dilaton Weyl background. Using the off-shell map given in \cite{Ozkan:2013uk}, a vector multiplet can be identified with fields in the gauged dilaton Weyl multiplet as
\begin{align}
	{\mathcal Y}^{ij} & \to \frac14 \rmi  \s^{-1} \bar\p^i \p^j - \frac{g}{2}  \s^{-1} L^{ij}\,, & \r & \to \s\,, \non\\
	 A_\m & \to C_\m \,, & \l^i & \to \p^i \,,
	\label{VecToDW2}
\end{align}	
	which gives rise to off-shell models that are purely expressed in terms of the fields of the (gauged) dilaton Weyl multiplet. By appropriately choosing primary composite linear multiplets, eq. \eqref{BF-Scomp} becomes the building block for constructing various curvature-squared invariants.
	
	In the superconformal approach, the off-shell formulation of minimal 5D	supergravity can be achieved by coupling the standard Weyl multiplet to two off-shell conformal compensators: a vector multiplet and a linear multiplet. Within this setup, supersymmetric completions of the Weyl tensor squared and Ricci scalar squared were constructed in \cite{Hanaki:2006pj} and \cite{Ozkan:2013nwa}, respectively.  The Weyl tensor-squared invariant is based on a composite linear multiplet comprised solely of standard Weyl multiplet fields. To construct the Ricci scalar-squared invariant, one starts by defining a composite vector multiplet in terms of a linear multiplet. This composite vector multiplet is then substituted into the vector multiplet action obtained using \eqref{BF-Scomp}.
	
	While the Weyl tensor-squared and Ricci scalar-squared actions based on the dilaton Weyl multiplet were presented as ungauged models \cite{Ozkan:2013nwa}, they can simply be gauged by using the maps \eqref{dilaton weyl basis} and \eqref{VecToDW2}. At this point, it is worthwhile to mention that the on-shell results for these gauged actions differ from those presented by \cite{Liu:2022sew}. The reason is that Ref. \cite{Liu:2022sew} assumes that the map between the standard Weyl and the dilaton Weyl multiplet is not modified in the gauged case. However, as shown in \cite{Coomans:2012cf} and presented as in eq. \eqref{dilaton weyl basis}, these expressions are indeed deformed. For completeness, we present the off-shell gauged results in the Supplemental Material. 
 
 The third invariant necessary to obtain all the curvature-squared models in five dimensions was constructed as the Riemann-squared invariant in the ungauged dilaton Weyl basis in \cite{Bergshoeff:2011xn}. However, this model does not refer to the standard Weyl multiplet; hence, the prescription to obtain gauged models cannot be applied. Furthermore, the construction methodology cannot be extended to the gauged dilaton Weyl multiplet. 
 
 Alternatively, a third independent, locally superconformal invariant
	containing the Ricci tensor-squared term can be constructed \cite{Butter:2014xxa,Gold:2023dfe}, which provides the correct basis to study gauged curvature-squared supergravity, as we shall discuss momentarily.  In this case, the lowest component of the composite linear multiplet is given by the field $L^{ij}_{\rm Log}$. This is obtained by making use of the standard Weyl multiplet and by acting with six $Q$-supersymmetry transformations on the field $\log{\rho}$, with $\rho$ being the lowest component of a compensating vector multiplet \cite{Note3}. The rest of the composite ``Log multiplet'' is then obtained by acting with up to two more $Q$-supersymmetry transformations on $L^{i j}_{{\rm Log}}$.	Due to the complexity of computing up to eight supersymmetry transformations, the explicit form of the  Log multiplet, including all fermionic terms, has been obtained only recently with the aid of the \textit{Cadabra} software \cite{Cadabra-1,Cadabra-2}. These lengthy results will be published elsewhere \cite{GJSG-2023}, see also \cite{Gold:2023dfe}, and \cite{GJSGMY-2023} for the complete analysis of the gauged supergravity case.
Inserting the resulting composite multiplet into \eqref{BF-Scomp} yields the explicit form of a new ``Log invariant'' which will be presented in \cite{GJSG-2023} in the standard Weyl basis. Then, the Log invariant in the gauged dilaton Weyl background can be obtained by employing the map \eqref{dilaton weyl basis} and \eqref{VecToDW2}. For the purpose of this letter, it suffices to present its bosonic sector in the gauge 
	\bea
	\sigma=1~, \qquad b_{\m}=0~, \qquad \psi^{i} = 0~.
	\label{gauge}
	\eea
	The gauged Log  invariant in the dilaton Weyl background, which includes a Ricci-squared term, reads
\begin{widetext}
	\bea
e^{-1}\cL_{{\rm Log}} 
  &=& 
	- \ft{1}{6} R_{{a} {b}} R^{{a} {b}}
        +\ft{1}{24}R^2 
	+\ft{1}{6}R^{{a} {b}}  G^{2}_{{a} {b}} + \ft{1}{3}R H_{{a} {b}} G^{{a} {b}}
	-\ft{4}{3}R_{{a} {b}}  H^{{a} {c}} G^{{b}}{}_{{c}} 
	-\ft{1}{3}R H^2
        - \ft{1}{12} \epsilon^{{a} {b} {c} {d} {e}} C_{{a}} V_{{b} {c}}\,^{i j} V_{{d} {e} i j} 
 \non\\
	&&+\ft{1}{6} V^{{a} {b} i j} V_{{a} {b} i j}
	- 2 ({{H}}^2)^2  +\ft{16}{3}H^{2}_{{a} {b}}  H^{{a}{c}} G^{b}{}_{c} - \ft{4}{3}H^2 H_{{a} {b}} G^{{a} {b}}
        +\ft{2}{3} H_{{a}{b}} H_{{c} {d}} \big( G^{{a} {b}} G^{{c} {d}} -2  G^{{a} {c}} G^{{b} {d}} \big)
 \non\\
	&&
	+\ft{2}{3}H^2 G^2
	-\ft{4}{3} H^{{2}{a} {b}}G^{2}_{{a} {b}}  - \ft{1}{3}H_{{a} {b}} G^{{a}{b}} G^2
	+G^{2}_{{a} {b}} H^{a c} G^{b}{}_{c}  - \ft{1}{48}(G^2)^2 -\ft{1}{24} G^4 
	-\ft{1}{6} \nabla_{{c}}G^{{a} {c}}  \nabla^{{b}}G_{{a} {b}}
 \non\\
	&&
	+2 \nabla_{{a}}H_{{b} {c}}   \nabla^{[{a}}H^{{b} {c}]}+ \ft{1}{48}\epsilon^{{a} {b} {c} {d} {e}} \nabla^{{f}}G_{{e} f}  (4 H_{{a} {b}} - G_{ {a} {b}}) (4 H_{{c} {d}} - G_{ {c} {d}}) 
 \non\\
	&& + \ft{g}{6}  \Big(   R N  -4 N H_{{a} {b}} G^{{a} {b}} - 2 N G^2 +  V_{{a} {b}}\,^{i j} L_{ij} ( G^{{a} {b}} + 4 H^{{a} {b}} )
	+12 N H^2 - 6 \nabla^{{a}} \nabla_{{a}}N 
	\Big)\non\\
	&&-  \ft{g^2}{24} \Big( 2 R L^2 -  L^2  (G^{2} -4 G^{{a} {b}}H_{{a} {b}} - 24  H^2 )  +4 N^2 +6 \nabla^{{a}}L^{ij}\nabla_{{a}}L_{ij}  \Big) +\ft{2}{3}N {L}^{2} {g}^{3}+\ft{5}{24}{L}^{4} {g}^{4}~,
	\label{log-gauged}
	\eea
	\end{widetext}
	where $V_{ab}{}^{ij} = 2 \partial_{[a} V_{b]}{}^{ij} - 2 V_{[a}{}^{k(i} V_{b]k}{}^{j)}$.  Furthermore, we have used the following notations: $H^{ab}=- \frac{1}{12}\epsilon^{abcde}H_{cde}$, $H^2 = H^{ab}H_{ab}$, $G^2 = G^{ab}G_{ab}$, $H^2_{ab}:=H_{a}{}^cH_{b c}$, $G^2_{ab}:=G_{a}{}^c G_{b c}$, $G^4 = G^{2 ab}G^{2}_{a b}$, and $H^4 = H^{2 ab}H^{2}_{a b}$.
 
 Note that the supersymmetric Riemann-squared invariant can be  obtained by taking the following linear combination of the ungauged Weyl-squared invariant presented in \cite{Bergshoeff:2011xn} and setting $g= 0$ in the Log invariant \eqref{log-gauged}
	\bea
 \cL_{{\rm Riem}^2} = \cL_{\rm Weyl^2} + 2 \cL_{\rm Log }|_{g=0}~.
	\eea
 The resulting action is identical to the one presented in \cite{Bergshoeff:2011xn} up to total derivatives.

\textit{Going on shell and dual CFT.}---Now	let us study a certain linear combination of the Einstein-Hilbert and all three curvature-squared invariants
	\be
	(16\pi G){\cal L}_{2\partial+4\partial}={\cal L}_{\rm EH}+\l_1 {\cal L}_{\rm Weyl^2}
 +\l_2 {\cal L}_{\rm Log}+\l_3{\cal L}_{R^2}\ .
	\label{totL}
	\ee
	where $G$ is Newton's constant and all the invariants are given in the gauged dilaton Weyl multiplet background. ${\cal L}_{\rm Weyl^2} $ and  ${\cal L}_{R^2}$ respectively denote the Weyl tensor squared and Ricci scalar squared actions which are obtained by employing the maps  \eqref{dilaton weyl basis} and \eqref{VecToDW2} in the standard Weyl multiplet results of \cite{Ozkan:2013nwa}.  Their explicit form is not crucial here, but they are given in the Supplemental Material for the reader's convenience.  The two-derivative Lagrangian ${\cal L}_{\rm EH}$ is obtained by using the linear multiplet action in the standard Weyl multiplet basis \cite{Coomans:2012cf,Ozkan:2013uk} and the sequential use of the maps \eqref{dilaton weyl basis} and \eqref{VecToDW2}.  Note that, in this section, we rescaled the Lagrangians such that the coefficient of their leading curvature-squared term is normalized to unity. To go on-shell, 	we fix the gauge according to \eqref{gauge} and break SU(2)   down to U(1)  by choosing
	\begin{align}
 L_{ij}& =\ft1{\sqrt{2}}\delta_{ij}L \,, & V_{a}^{ij} &=V_{a}^{'ij}+\ft12\delta^{ij}V_{a}\ .	
	\end{align}
	Consequently, the two-derivative  Lagrangian becomes
	\bea
	&&e^{-1}{\cal L}_{\rm EH}=L(R-\ft12G_{ab}G^{ab}+4H_{ab}H^{ab}+2V_{a}^{'ij}V^{'a}_{ij})
	\nonumber\\
	&&+L^{-1}\partial_{a} L\partial^{a} L-2L^{-1}E_{a}E^{a}-2\sqrt{2}E_a V^a -2N^2L^{-1}
	\nonumber\\
	&& -4g C_{a}E^{a}-2gNL-4gN-\ft12 g^2L^3+2g^2L^2\ .
	\eea
	From the total Lagrangian \eqref{totL}, several auxiliary fields can be solved from their field equations up to ${\cal O}(\l_i)$  
	\bea
	N&=&-\ft{1}2 gL(2+L)+{\cal O}(\l_i)\ ,
	\nn\\
	\quad E_{a}&=&{\cal O}(\l_i)\,,\quad V^{'ij}_{a}={\cal O}(\l_i)\ .
	\label{auxsol}
	\eea
 To arrive at the five-dimensional gauged minimal supergravity, we first dualize $B_{\m\n}$ to a new 1-form gauge field $\widetilde{C}_{\m}$ following the procedure in \cite{Coomans:2012cf,Ozkan:2013uk}. We then truncate the model consistently by imposing
	\be
	L=1+{\cal O}(\l_i)\,,\quad \widetilde{C}_{a}=C_{a}+{\cal O}(\l_i)\ .
	\label{trun}
	\ee
 Following \eqref{trun}, the field equation of $E_{abc}$ now implies
 \be
 V_{a}=-\ft{3}{\sqrt 2}gC_{a}+{\cal O}(\l_i)\ .
 \label{auxsol-2}
 \ee
 Plugging \eqref{auxsol}--\eqref{auxsol-2} back to the total Lagrangian \eqref{totL},  one obtains the on-shell theory up to first order in $\l_i$.
 It is important to note that in the procedure outlined above, the ${\cal O}(\l_i)$ terms arising from substituting \eqref{auxsol}--\eqref{auxsol-2} to the two-derivative action either vanish (proportional to the leading order equations of motion of auxiliary fields) or can be removed by field redefinitions \cite{Argyres:2003tg}. 
 To recover the standard convention of minimal supergravity, we rescale the graviphoton and the U(1) coupling according to $C_{a}\rightarrow\ft1{\sqrt 3}C_{a},\,g\rightarrow\sqrt2 g$.
 To conclude, following \cite{Myers:2009ij}, the resulting Lagrangian can be further simplified by redefining the metric and the U(1) gauge field.  Eventually, the on-shell model is recast in the form below
	\bea
	(16\pi G)e^{-1}{\cal L}_{2\partial+4\partial}
	=c_0 R+12c_1 g^2 -\ft{1}4 c_2 G_{ab} G^{ab}
	\nn\\
 +\ft{1}{12\sqrt3}c_3\epsilon^{abcde}C_{a}G_{bc} G_{de}
 +\l_1 {\cal L}_{\rm GB}|_{\rm onshell}\ ,
	\label{toton}
	\eea
	where the various coefficients are
	\bea
	c_0&=&1+(\ft{28}{3}\l_1 -20\l_2 -4\l_3) g^2\, ,
	\nn\\
	c_1&=&1+(\ft{50}{9}\l_1 -\ft{28}{3}\l_2+\ft{52}3\l_3) g^2\ ,
	\nn\\
	c_2&=&1+(\ft{64}9\l_1 -\ft{92}3\l_2 -\ft{76}3\l_3) g^2\ ,
	\nn\\
	c_3&=& 1-12  (\l_1+3\l_2+3\l_3)g^2\ ,
	\eea
	and the on-shell Gauss-Bonnet invariant is given by
	\bea
{\cal L}_{\rm GB}|_{\rm onshell} 
= 
R_{abcd}R^{abcd}
-4R_{ab}R^{ab}+R^2
+\ft18 G^4
\nn\\
 -\ft12 W_{abcd}G^{ab}G^{cd}
+\ft1{2\sqrt3}\epsilon^{abcde}C_{a}R_{bc}{}^{fg}R_{defg}
 ~,
	\eea
	where $W_{abcd}$ is the Weyl tensor. This on-shell action is consistent with the generic result presented in \cite{Cassani:2022lrk} for the proper choice of parameters. Based on the on-shell model \eqref{toton}	we find that the AdS$_5$ radius receives corrections from the higher-derivative terms and is given by 
	\be
	\ell=g^{-1}(1+\ft{8}{9}  g^2 \l_1-\ft{16}{3} g^2\l_2-\ft{32}{3} g^2 \l_3)\ .
	\label{ads5l}
	\ee
The effective Newton's constant from Eq. \eqref{toton} is then 
	\be
	G_{\rm eff}=G + G (-\ft{28}{3}  \l_1+20 \l_2+4  \l_3)g^2\ .
	\label{effg}
	\ee
 The AdS$_5$ vacuum preserves maximal eight supercharges \cite{Gutowski:2004ez,Gutowski:2004yv} and the dual field theory should be a $D=4,\, \mathcal{N}=1$ CFT. Utilizing \eqref{ads5l} and \eqref{effg} the $a$ and $c$ Weyl anomaly coefficients of the dual CFT  can be obtained via the standard  holographic renormalization procedure \cite{Henningson:1998gx, Fukuma:2001uf}
	\bea
	a&=&\frac{\pi }{8 g^3 G}-\frac{9 \pi  (\l_2+\l_3)}{2 g G}
 ~,
	\nn\\
	c&=&\frac{\pi }{8 g^3 G}+\frac{\pi  (2\l_1-9\l_2-9\l_3)}{2 g G}
 ~,
	\eea
	using which one finds the results above are  consistent with $R$-symmetry anomaly whose coefficients are related to those of the Weyl anomaly via \cite{Anselmi:1997am,Cassani:2013dba}
	\be
	5a-3c=\frac{\pi c_3}{4g^3 G}
 ~,\quad 
 a-c=-\frac{\pi  \l_1}{g G}\ .
	\ee

\textit{Conclusions and outlook.}---In this letter, we provide the correct and complete basis to study curvature-squared gauged supergravity in five dimensions. Based on the new results, we successfully computed the anomaly coefficients governing dual four-dimensional SCFTs. As four-dimensional SCFTs are characterized by two anomaly coefficients, one would naturally anticipate the emergence of only two independent linear combinations among the three four-derivative couplings. Indeed, our analysis confirms this expectation, with $\lambda_2$ and $\lambda_3$ consistently appearing together in the anomaly coefficients as the combination $\lambda_2+\lambda_3$. However, when examining the on-shell action \eqref{toton}, it becomes evident that $\lambda_2$ and $\lambda_3$ do not share this combination. This suggests that while two curvature-squared invariants may suffice for calculating BPS quantities \cite{Baggio:2014hua,Bobev:2021qxx,Bobev:2022bjm,Cassani:2022lrk}, the computation of generic physical parameters may require the incorporation of all three invariants.
 
  We can also generalize the results by coupling multiple vector multiplets which also enjoy an off-shell formulation. Given the simple form the 6D ungauged Gauss-Bonnet invariant \cite{Liu:2013dna,Novak:2017wqc,Butter:2018wss} and the relation between dilaton Weyl multiplets in these two dimensions \cite{Bergshoeff:2011xn}, it may be feasible to reformulate the 5D Gauss-Bonnet invariant into a more elegant expression that facilitates the construction of intriguing solutions. For instance, the non-existence of supersymmetric AdS$_5$ black ring solutions in the two-derivative theory \cite{Grover:2013hja} raises the intriguing question of whether this situation changes in the presence of higher-derivative interactions.  Our new invariants enable the computation of corrections to the entropy of $\ft1{16}$-BPS black holes \cite{Gutowski:2004ez,Gutowski:2004yv,Cvetic:2004ny,Chong:2005hr,Kunduri:2006ek}, thereby extending the precision test of black hole microstate counting to the next to leading order. It is also interesting to extend the recently proposed equivariant localization \cite{BenettiGenolini:2023ndb, BenettiGenolini:2023kxp} beyond the leading two-derivative cases.

\textit{Acknowledgements.}---The work of G.G., J.H., S.K., and G.T.-M.	was supported by the Australian Research Council (ARC)	Future Fellowship FT180100353, and by the Capacity Building Package of the University	of Queensland.
G.G. and S.K. are supported by postgraduate scholarships at the University of Queensland.
	 M.O. acknowledges the support by the Outstanding Young Scientist Award of the Turkish Academy of Sciences (TUBA-GEBIP). 
  The work of Y.P. is supported by National Natural Science Foundation of China (NSFC) under grant No. 12175164 and the National Key Research and Development Program under grant No. 2022YFE0134300.

 \clearpage

\appendix{}

\begin{center}\textbf{SUPPLEMENTAL MATERIAL\\
(APPENDICES)}\end{center}

In the Supplemental Material to our letter, we present the bosonic sectors of the off-shell Weyl-squared and scalar curvature-squared invariants in the gauged dilaton Weyl background. These results can be obtained by starting from the invariants written in a standard Weyl multiplet background, then applying the map of fields \eqref{dilaton weyl basis} which defines the gauged dilaton Weyl multiplet.

\section{A. The gauged Weyl-squared invariant}
The bosonic sector of the gauged Weyl-squared invariant in the gauged dilaton Weyl background, eq.~\eqref{dilaton weyl basis}, and in the gauge \eqref{gauge}, is given by:
\begin{widetext}
\bea
    e^{-1}\cL_{{\rm Weyl}^2} 
    &=&  
    - \ft{1}{8} \epsilon^{{a} {b} {c} {d} {e}} C_{{a}} R_{{b} {c} f g} R_{{d} {e}}{}^{f g} 
    + \ft{1}{6} \epsilon^{{a} {b} {c} {d} {e}} C_{{a}} V_{{b} {c}}\,^{i j} V_{{d} {e} i j} 
    +\ft{2}{3} V^{{a} {b} i j} V_{{a} {b} i j} 
    - \ft{1}{4}R_{{a} {b} {c} {d}} R^{{a} {b} {c} {d}} 
    +\ft{1}{3}R_{{a} {b}} R^{{a} {b}}  
    - \ft{1}{12}{R}^{2}
    \non \\
    &&  
    + \ft{1}{3}R_{{a} {b} {c} {d}} ({{G}}^{{a} {b}} G^{{c} {d}} 
    - 2 H^{{a} {b}} H^{{c} {d}} -3 {{H}}^{{a} {b}} {{G}}^{{c} {d}}) 
    - \ft{4}{3} R_{{a} {b}} {{H}}^{{a} {c}} G^{{b}}{}_{{c}}  
    +\ft{16}{3}R^{{a} {b}} {{H}}^{2}_{{a} {b}} 
    + \ft{1}{3}R {{H}}_{{a} {b}} G^{{a} {b}} 
     -\ft{4}{3}R H^2
    \non \\
    &&     
    - 4(H^2)^2 
    -8H^4
    -\ft{16}{3} H^2 H_{{c} {d}} G^{{c} {d}} 
    -\ft{40}{3}H^{2}_{{a} {d}} H^{a}{}_{ {c}}  G^{{c} {d}}
    +\ft{8}{3}H^2 G^2
    +\ft{2}{3} H_{{a} {b}} H_{{c} {d}} G^{{a} {b}} G^{{c} {d}}    
    + \ft{1}{12}(G^2)^2 
    \non \\
    &&
    - \ft{16}{3}H^{2}_{{a} {b}}  G^{2 {a} {b}}  
    - \ft{4}{3} H_{{a} {b}} H_{{c} {d}} G^{{a} {c}} G^{{b} {d}}
    - \ft{1}{3}H_{{a} {b}} G^{{a} {b}} G^{2}  
    +2 G^{2 {a} {b}} G^{{c}}{}_{{b}} H_{{c} {a}}   
    -\ft{1}{3} (\nabla^{{a}}{G_{{b} {c}}}) \nabla_{{a}}{G^{{b} {c}}} 
    \non\\
    &&
     +\ft{8}{3} (\nabla^{{a}} H_{{b} {c}} )\nabla_{{a}} H^{{b} {c}} 
    - \ft{1}{2}G^{4}  
    +\ft{4}{3}\epsilon^{{a} {b} {c} {d} {e}} H_{{a} {b}} H_{{c} {d}} \nabla^{{f}}{G_{{e} {f}}}
    - 2 \epsilon^{{a} {b} {c} {d} {e}}  H_{{b} {f}} (\nabla_{{a}}H_{{c}}{}^{{f}}) G_{{d} {e}}
     \non \\
    &&
    - \ft{2}{3}\epsilon^{{a} {b} {c} {d} {e}} H_{{a} {b}} (\nabla^{{f}}{G_{{c} {f}}} ) G_{{d} {e}}
    -\ft{1}{24}\epsilon^{{a} {b} {c} {d} {e}} (\nabla^{{f}}{G_{{a} {f}}}) G_{{b} {c}} G_{{d} {e}} 
    \non \\
    && -g \left( -\ft{4}{3} N H_{a b} G^{a b}  + \ft{4}{3} N G^{2} -8 N H^{2}
     +\ft{4}{3} V_{a b}{}^{ij} L_{ij} H^{a b} - \ft{2}{3}V_{a b}{}^{ij} L_{ij} G^{a b} - \ft{2}{3} R N \right)
    \non\\
    &&+ g^2 \left( \ft{1}{3}L^2 H^{a b} G_{a b} -\ft{1}{3}L^2 G^{2} +2 L^2 H^{2} -\ft{8}{3}N^2 + \ft{1}{6} R L^2\right) -\ft{4}{3} g^3 N L^2 -\ft{1}{6} g^4 L^4~,
\eea
\end{widetext}
where $H^4:=H^2{}^{ab}H^2_{ab}$ and other combinations of product of fields are defined after eq. \eqref{log-gauged}. 

\section{B. The gauged scalar curvature-squared invariant}
The bosonic sector of the gauged scalar curvature-squared invariant in the gauged dilaton Weyl background, eq.~\eqref{dilaton weyl basis}, and in the gauge \eqref{gauge}, is given by:
\begin{widetext}
    \bsubeq
\bea\label{R2-1}
     e^{-1}  \cL_{R^2} &=&  \mathbf{Y}^{i j} \mathbf{Y}_{i j} -  2\de^{{a}}(N L^{-1})\de_{{a}}(N L^{-1})  -\ft{1}{8}  \e_{{a} {b} {c} {d} {e}}C^{{a}}\mathbf{G}^{{b} {c}}\mathbf{G}^{{d} {e}} + N L^{-1} G^{{a} {b}}  \mathbf{G}_{{a} {b}}  -   N^2 L^{-2} G^{a b} G_{a b}    \non \\
    &&+ 4N^2 L^{-2} H^{a b}G_{a b}     - \ft{1}{4} \mathbf{G}^{{a} {b}}\mathbf{G}_{{a} {b}}  -4 N L^{-1}  H_{a b}  \mathbf{G}^{{a} {b}} - 4 g   N^3 L^{-2}  + \ft{1}{16} g^4   L^4 \non \\
   &&
    +  g^2  \Big(    \ft{1}{4} L^{ij} \de^{a} \de_{a}{L_{ij}} 
    -  \ft{1}{4}R L^2
-  H^{2} L^2
 +  \ft{1}{8} G^{2} L^2 - \ft{5}{2}  N^2 
 -  \ft{1}{2} E^{a}E_{a}   - \ft{1}{2} \de^{a}{L} \de_{a}{L}
\Big) ~,
     \label{R2-eq}
\eea
where,
\bea
 \mathbf{G}_{{a} {b}} 
&=&   4 \nabla_{[a} ( L^{-1} E_{b]}) + 2 L^{-1} L_{ij}(V_{ab}{}^{ij})
-2 L^{-3}  L_{ij}  \big(\nabla_{[a} L^{ik} \big) \nabla_{b]} L_k{}^{j}~,
\non\\ 
    \mathbf{Y}^{ij}{} 
&=&   \ft{1}{4} L^{-1} \left( 4 \nabla^{a} \nabla_{a}{L^{ij}} - 2 R {L^{ij}}
- 8
 H^{2} {L^{ij}}
 +   G^{2} {L^{ij}}
\right)
\non\\
&&+  L^{-3} \left( - N^2 {L^{ij}} -  E^{a}E_{a} {L^{ij}} - 2 E^{a}L^{k(i} \nabla_{a} L_k{}^{j)}- {L_{kl}} \nabla^{a}  {L^{k(i}} \nabla_{a} {L^{j)l}}
\right) \ .
\eea
\esubeq
\end{widetext}

\end{document}